\documentclass[prl,twocolumn,showpacs,preprintnumbers,amsmath,amssymb]{revtex4}
\usepackage{graphicx}% Include figure files
\usepackage{dcolumn}% Align table columns on decimal point
\usepackage{bm}% bold math
\usepackage{psfig}
\newcommand \be{\begin{eqnarray}}
\newcommand \ee{\end{eqnarray}}

\begin{document}
\title{Two-particle binding energy of interacting Bose gases}
\author{K. Morawetz$^{1,2}$, M. M{\"a}nnel$^{1}$, M. Schreiber$ ^1$ and P. Lipavsk\'y$^{3,4}$}
\affiliation{$^1$Institute of Physics, Chemnitz University of Technology, 
09107 Chemnitz, Germany}
\affiliation{$^2$Max-Planck-Institute for the Physics of Complex
Systems, N{\"o}thnitzer Str. 38, 01187 Dresden, Germany}
\affiliation{$^3$Faculty of Mathematics and Physics, Charles University, 
Ke Karlovu 3, 12116 Prague 2, Czech Republic}
\affiliation{$^4$Institute of Physics, Academy of Sciences, 
Cukrovarnick\'a 10, 16253 Prague 6, Czech Republic}

\begin{abstract}
The pole of the two-particle T-matrix including the influence of the surrounding medium is analyzed for an interacting Bose gas. The phase diagram of the Bose-Einstein condensation (BEC) depending on the temperature, density, scattering length, and momentum is derived from this pole. The critical momentum for the occurrence of superfluidity is obtained in this way. As a new observation a two-particle binding energy is reported intimately connected with the occurrence of the BEC. It is suggested that this might have cosmological consequences on the dark energy problem. 
\end{abstract}
\date{\today}
\maketitle

There are interesting scenarios which attribute the unsolved problem of dark matter and energy in the universe to interacting bosonic degrees of freedom \cite{RT00}. One of the experimental hints is the excess of diffusive gamma rays which is attributed to annihilation of dark matter in our galaxy \cite{MP07}.  The observed halo structures and the observed matter distribution which is not in agreement with the gravitational law might be accounted for by almost massless (pseudo-)scalars such that the corresponding Klein-Gordon equation leads to BEC of astronomic size \cite{MP07}. Scalar field models can in principle describe both dark energy and dark matter \cite{BKPU03,MP07}. Such point-like cold particles are postulated to explain the structure of self-gravitating objects extending the standard model \cite{BG06,BG06a}. Interactions between postulated scalars and baryons embedded in the condensate can also explain the anomalies in the cosmic microwave background \cite{FG04}.

The idea to identify the dark energy with the BEC of some boson field remains under debate \cite{FM06}. In this model a mechanism is proposed that the condensate collapses to black holes and other localized objects rapidly with new condensation starting. In this way the universe is thought to be filled with many such black holes and localized objects entering the inflationary regime. The dark matter is understood as formed by collapses of the dark energy.  
In this letter we show that the appearance of BEC in an interacting Bose gas is everytime connected with a binding energy. Since in the BEC scenarios described, the energy balance is counted without this binding energy, our observation might have an impact on the above mentioned models. Especially it might contribute to the missing dark energy balance.

The BEC phase transition occurs since the free energy of the condensed phase is lower than the free energy of the normal phase such that it is sometimes viewed as a first-order phase transition \cite{H85}. 
This is connected with interactions between 
the particles. BEC appears already in the ideal Bose gas since one takes into account at least Bose statistics as the correlation. Generally speaking, any interaction energy is a 
two-particle quantity. Therefore it is tempting to analyze the BEC in terms 
of two-particle properties. We will perform this task with the help of the 
two-particle T-matrix taking into account the influence of the surrounding 
medium. We will derive the conditions for BEC from the pole of the T-matrix.

Interacting Bose systems have recently gained large attention. This is partially due to the success of tabletop experiments of interacting Bose gases in traps  \cite{AWMEC95,DMADDKK95}. Mainly the change of the critical temperature is in the center of interest \cite{A04,BBHLV01}. For bulk Bose systems the critical temperature rises linearly with interaction \cite{HGL99,BR02,KNP04,PST01,HLK04} in agreement with the Monte Carlo data \cite{HK99,AM01,KPS01}. For stronger interactions a nonlinear behavior appears with a maximal critical temperature and a maximal interaction up to which Bose-Einstein condensation (BEC) is possible \cite{KSP04a,MMS07}. The appearance of bound states have been not considered on the same footing. We will fill this gap with the present paper.

The bosonic two-particle T-matrix for contact interaction in a bosonic medium depends on the frequency $\omega/\hbar$ and the center-of-mass momentum $Q$ relative to the surrounding medium \cite{MMS07}
\be
T(Q,\omega) = - {4 \pi \hbar^2 a_0 \over m}{1 \over 1 + J(Q,\omega)}
\label{JJci}
\ee
\be
J(\!Q,\omega) \!\!\!&=&\!\!\!{4 \pi a_0 \hbar^2\over m} \!\sum\limits_q \! \left (\!{m\over q^2}\!+\! {1\!+\!f(\epsilon_{\frac Q 2 \!-\!q}\!-\!\mu)\!+\!f(\epsilon_{\frac Q 2 \!+\!q}\!-\!\mu)\over \omega\!-\!\epsilon_{\frac Q 2 \!-\!q}\!-\!\epsilon_{\frac Q 2 \!+\!q}\!+\!i \eta}\!\right )
\nonumber\\&&
\label{J}
\ee
with the quasiparticle dispersion $\epsilon_p=p^2/2m$, the Bose distribution $f(x)=1/[\exp{(x/T)}-1]$, and the free scattering length $a_0$ without medium effects. Without loss of generality we restrict our investigation here to the nonrelativistic case which can be generalized to relativistic  dispersions.

\begin{figure}
\psfig{file=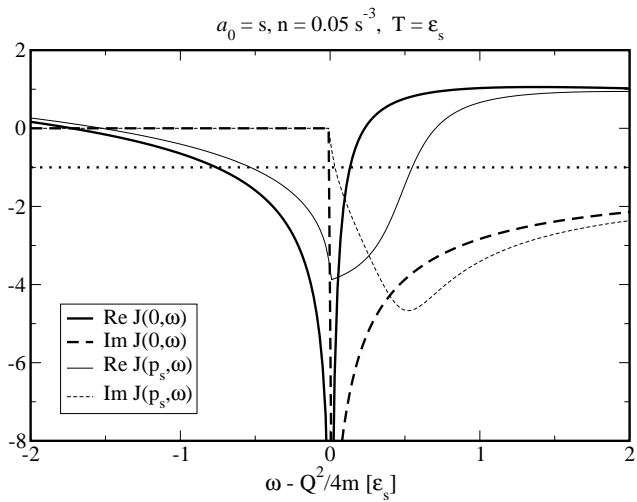,width=9cm}
\caption{The real and imaginary part of $J(Q,\omega)$ versus energy relative to the center-of-mass energy. The length scale $s$ is fixed 
to the scattering length. The corresponding scales for energy and momentum are $\epsilon_s=\hbar^2/2ms^2$ and ${\rm p}_s= \hbar/s$.}
\label{jqs}
\end{figure}

We will now determine the pole of the T-matrix which provides information about the two-particle excitations in the system. In particular we will find the condition for the occurrence of BEC. In order to justify this approach we compare the results here with the condition for BEC found earlier. The medium-dependent scattering length $a$ diverges at the phase transition \cite{MMS07,SMR97} and is given by the expansion with respect to small momentum, $p=\hbar k$, of the scattering phase 
\be
\tan{\delta}=\left . {{\rm Im} T\over {\rm Re} T}\right |_{\scriptsize \begin{array}{l}Q=0
\cr \omega ={Q^2\over 4m}+{p^2\over m}
\end{array}} 
\approx a k+{\cal O}(k^2).
\label{adef}
\ee
For quadratic dispersion %and $Q=0$ 
we have
\be
{a\over a_0}={1+2 f(\epsilon_0-\mu)\over 1-{4 a_0 \over \pi \hbar}\int\limits_0^\infty d{\bar p} f(\epsilon_{\bar p}-\mu)}.
\label{aa0}
\ee
The medium-dependent scattering length $a$ diverges as soon as the
scattering length $a_0$ approaches the critical one 
\be
a_0^C = \lim_{a \to \infty} a_0 =\left({4 \over \pi \hbar} \int\limits_0^\infty d\bar p f(\epsilon_{\bar p}-\mu) \right)^{-1}
\label{SCa}
\ee
where the denominator of (\ref{aa0}) is zero.

Now let us investigate the pole condition of the T-matrix (\ref{JJci}), 
i.e., $J=-1$. In figure \ref{jqs} we plot the real and imaginary part of 
%(\ref{bcond})
(\ref{J}). We see that the imaginary part is finite for positive energies describing scattering excitations while for negative energies relative to the center-of-mass energy no scattering dissipation limits the lifetime of the excitation. The real part is sharply peaked with a minimum at the continuum threshold $\omega=Q^2/4m$. Therefore in order to have a pole one has to demand 
\be
{\rm Re} J\left(Q,{Q^2 \over 4 m}\right)\le-1.
\label{QC}
\ee
From this condition and inspecting (\ref{J}) we obtain in fact two conditions. We find that the real part of $J(Q,Q^2/4m)$ is a monotonically increasing function of the momentum $Q$. Therefore the first condition to have a pole reads
\be
{\rm Re} J(0,0)\le-1,
\label{SC}
\ee
or equivalently $a_0\ge a_0^C$
called scattering condition in the following.
The second condition, called momentum condition, requires that the center-of-mass 
momentum is smaller than a critical value
\be
Q \le Q_c
\label{qc}
\ee
determined by the equality sign in (\ref{QC}).

\begin{figure}[h]
\centerline{\parbox[t]{9cm}{
\parbox[t]{4.5cm}{
\psfig{file=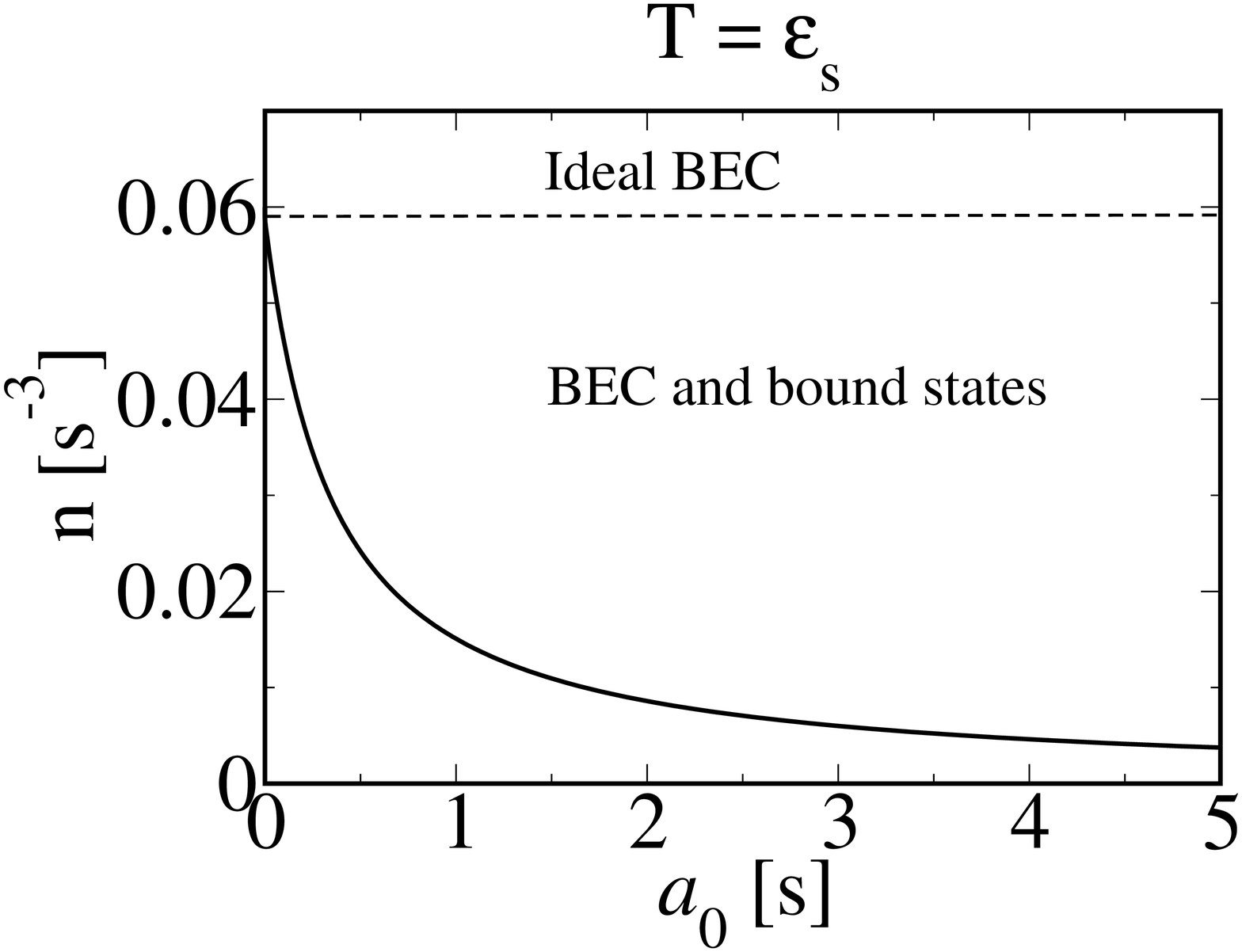,width=4.5cm}}
\hspace*{-2ex}
\parbox[t]{4.5cm}{
\psfig{file=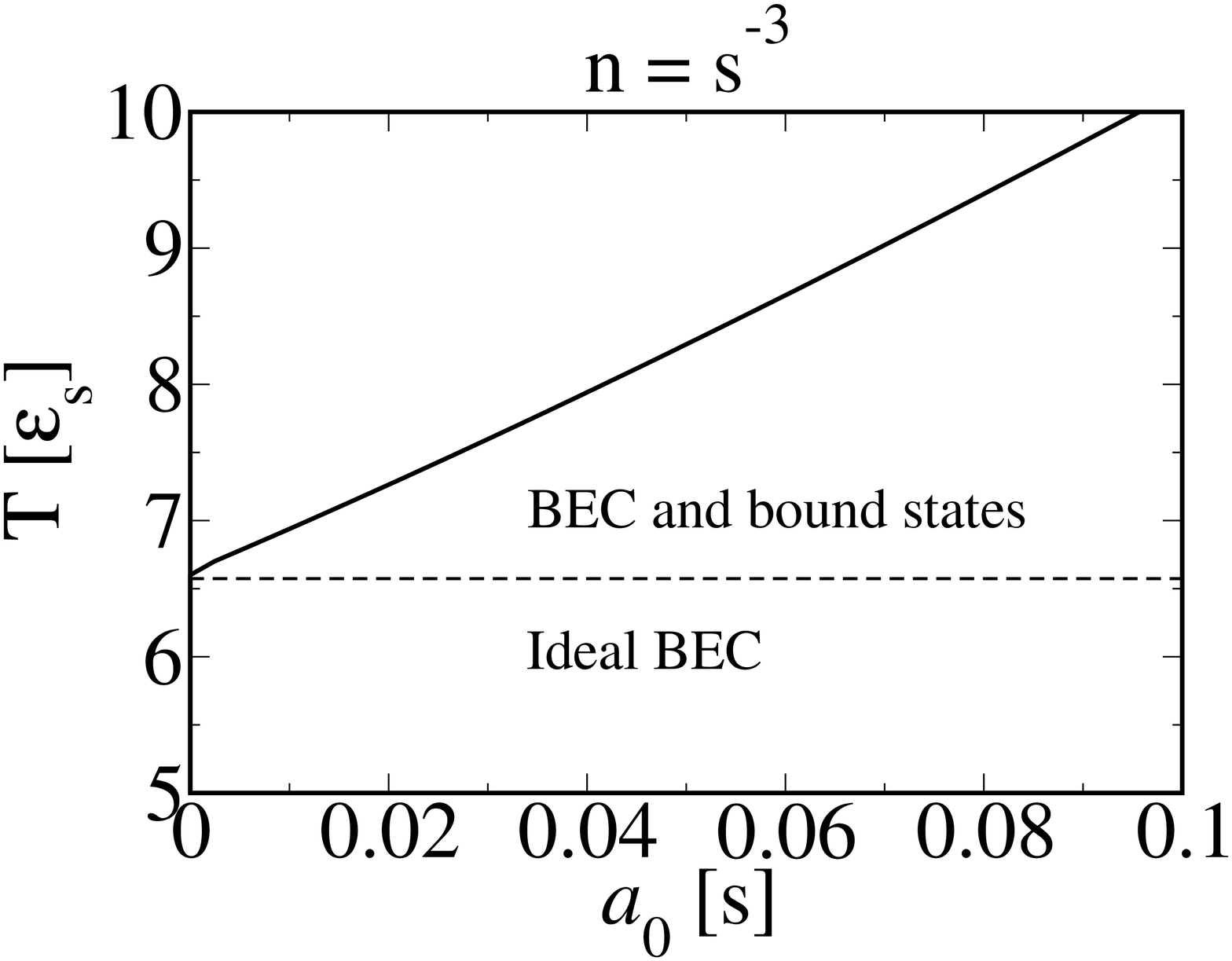,width=4.5cm}}
}}
\caption{The critical density for a fixed temperature (left) and critical temperature for
a fixed density (right) versus scattering length. The scale parameter $s$ is fixed by the temperature and density and determines the scales.}\label{tac}
\end{figure}

Using the scattering condition we are able to determine the phase diagram with respect to the density, temperature, and interaction length. In figure \ref{tac} we give this phase diagram as it is obtained from the T-matrix.
The critical temperature rises and the critical density decreases with increasing scattering length while for the noninteracting (ideal) case it remains constant. One should note that the T-matrix approximation is a weak-coupling approximation and for higher values of the scattering length the critical temperature is not increasing further but decreasing and reaching zero at a maximal critical scattering length \cite{MMS07}. 

\begin{figure}[h]
\centerline{\parbox[t]{9cm}{
\parbox[t]{4.5cm}{
\psfig{file=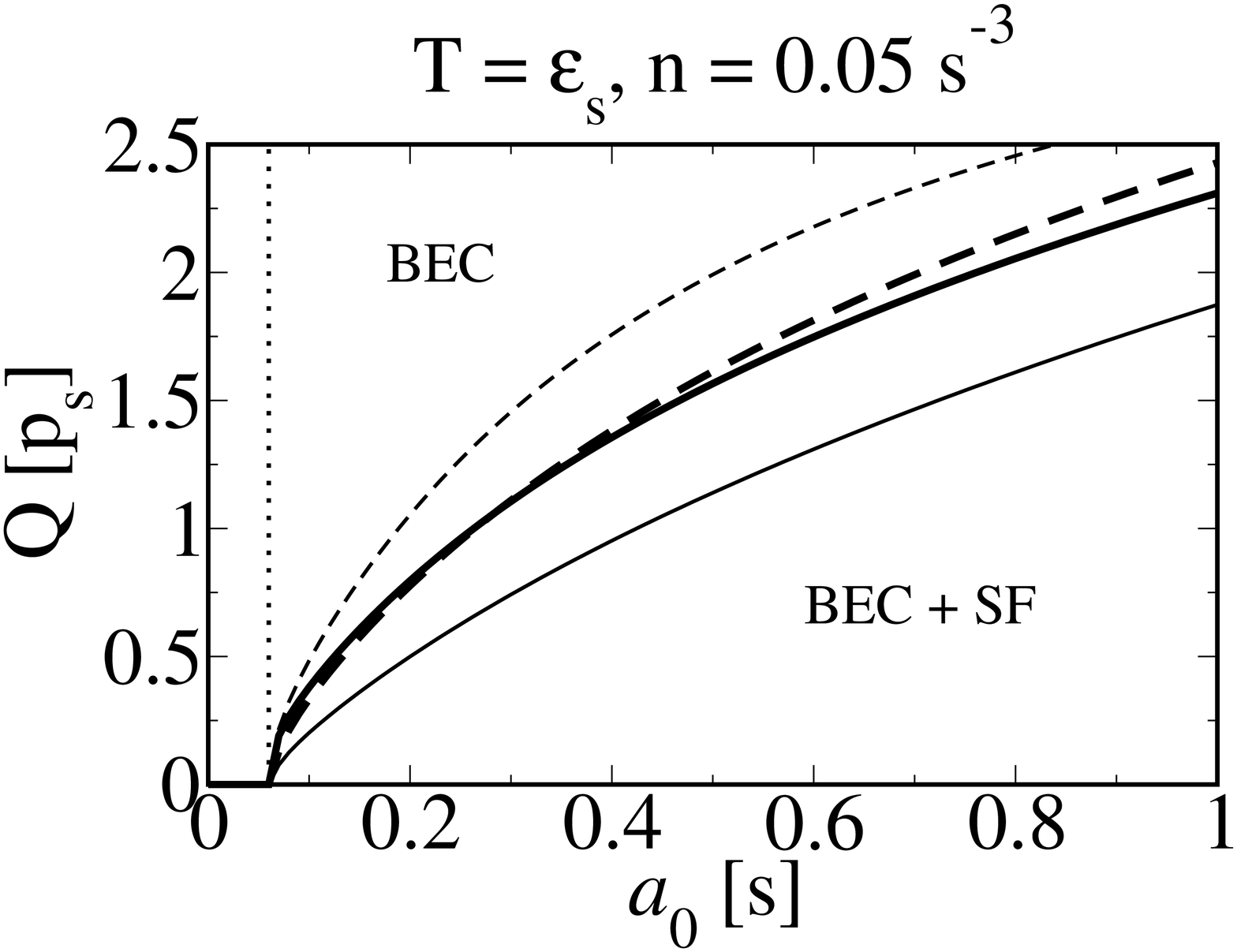,width=4.5cm}}
\hspace*{-2ex}
\parbox[t]{4.5cm}{
\psfig{file=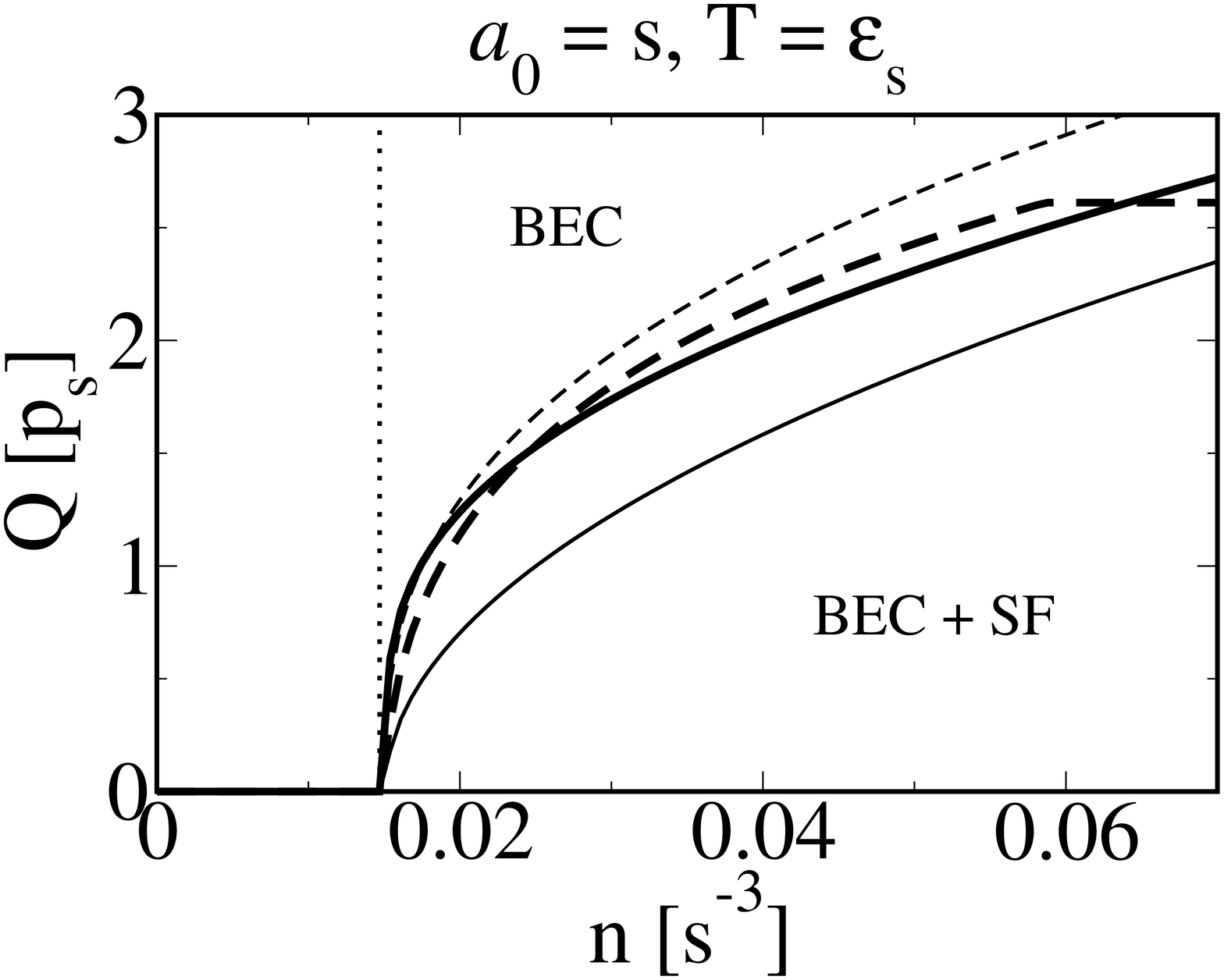,width=4.5cm}}
\vspace*{-3ex}
\parbox[t]{5cm}{
\psfig{file=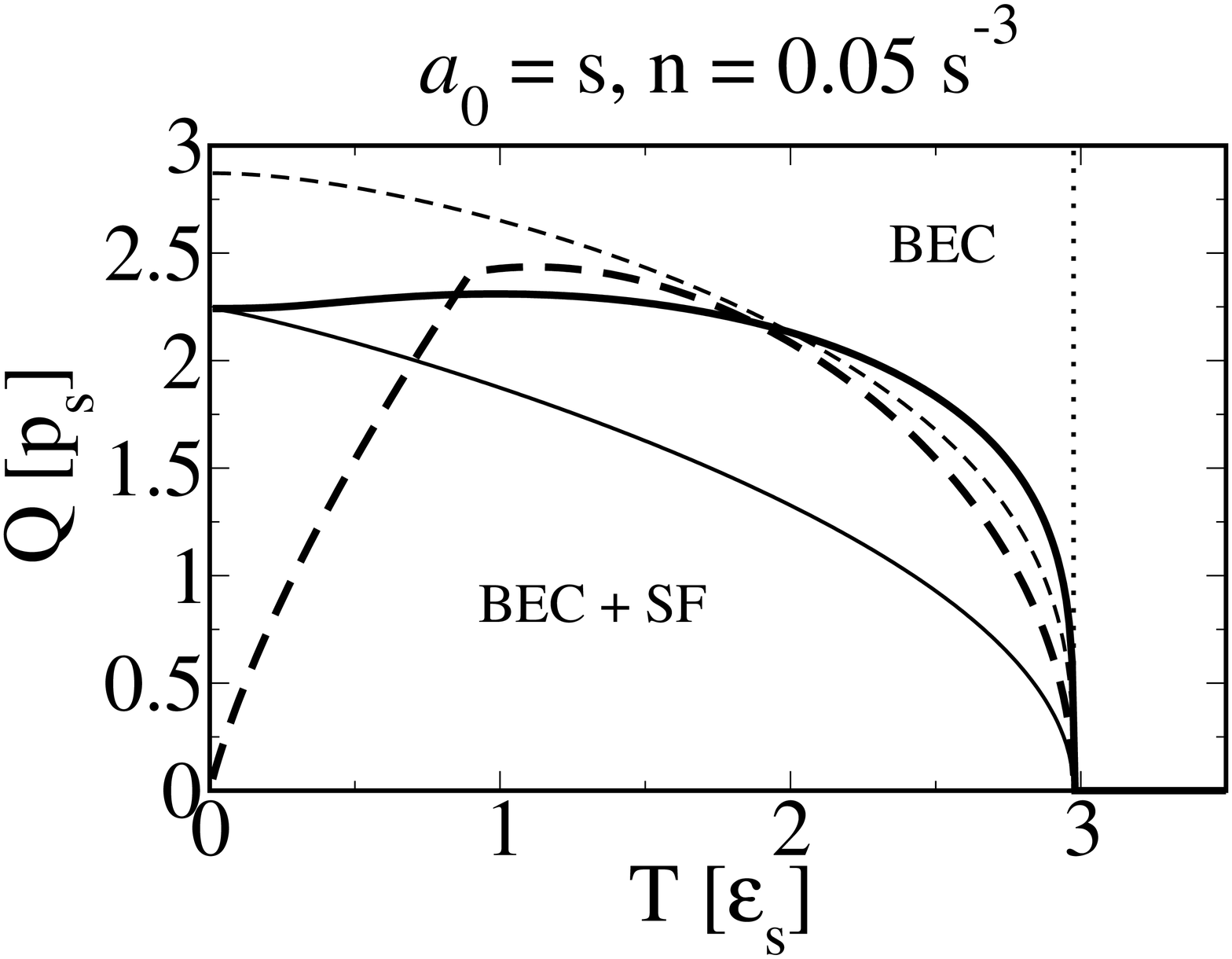,width=5cm}}
}}
\caption{The critical momentum versus scattering length for a fixed temperature and density (upper left), versus density for a fixed temperature and scattering length (upper right) and versus temperature for a fixed density and scattering length (below). The curves without condensate density (thick broken lines) are compared with the curves including the condensate (thick solid lines). The condensation starts or ends at the critical values obtained from the scattering condition (vertical dotted lines). Critical momenta according to (\ref{qc1}) (thin broken lines) and (\ref{qc2}) (thin solid line) are also included.%The dotted line separates the BEC region.
}\label{qct2}
\end{figure}

The momentum condition leads to an interesting dependence of the pole on the center-of-mass momentum. In figure \ref{qct2} we present the dependence of the critical momentum on the different parameters.
We interpret the occurrence of an upper critical momentum for the pole in the T-matrix as a finger print for superfluidity. Whether BEC occurs or not is seen in the center-of-mass system such that only the scattering condition (\ref{SC}) decides whether BEC occurs or not. The critical momentum according to the momentum condition (\ref{qc}) now separates the phase diagram additionally into an area with only BEC and an area with BEC and superfluidity. 

That this interpretation is justified
we can understand by a reasoning similar to the derivation of the Landau condition for superfluidity. We imagine a two-boson bound state dragged through a medium of ground-state bosons. The binding energy $E_B$ is obtained from the pole of the T-matrix for $Q=0$, neglecting the $Q$ dependence of $E_B$. In case of sufficiently high momentum, $Q=2 m |V|$, dissipation will happen such that the bound state will scatter with the Bose particles and will break off. If the momentum and energy of a single particle of the medium are changed to $p$ and $\epsilon=p^2/2m$ by the scattering, one can find that the momentum and energy transfered to the two-particle object are $\Delta P=-p$ and $\Delta E= pV-\epsilon$ \cite{V03}. If the bound state breaks off, the latter must account for $E_B$ and the center-of-mass energy $p^2/4m$. From the resulting balance, $|p| |V| \cos\theta-\epsilon\equiv E_B+p^2/4m$, we obtain a condition for such process to happen 
\be
Q&=&{2 m (\epsilon+E_B)+\frac{1}{2} p^2 \over |p| \cos\theta}%\ge {2 m (\epsilon+\Delta_B)+\frac{1}{2} p^2 \over |p|}
\nonumber\\
&\ge&{2m E_B\over |p|}+\frac{3}{2} |p|\ge 2\sqrt{3 m E_B}\equiv Q_c.
\label{qc1}
\ee
In other words for momenta lower than $Q_c$ such dissipative processes cannot occur and the medium behaves without dissipation, i.e., frictionless and therefore superfluid.

In figure~\ref{qct2} we plot the results for the critical momentum obtained from the momentum condition of the T-matrix with and without condensate.
The kinks at the critical values of the ideal Bose gas vanish when the condensate density is included in the following way. For homogeneous Bose systems one separates the condensate in the distribution function \cite{ST97} as
\be
f(\epsilon_p-\mu) = {1 \over e^{(\epsilon_p - \mu)/T} - 1} + (2\pi\hbar)^3 n_0 \delta(p).
\label{fp}
\ee
The density of the condensate $n_0$ can be determined by selfconsistent 
approaches \cite{G95,SG98}.
Here we use a simplified model which is sufficient for the present purpose. For densities 
smaller than the critical one the condensate density $n_0$ is zero. For densities 
larger than the critical one we keep the chemical potential fixed at the critical value $\mu^C$ given by (\ref{SCa}) with $a_0^C(\mu^C)=a_0$. Then we calculate the density of the non-condensate particles, from the first part of (\ref{fp}) at this critical chemical potential. The difference between the total density and the normal density 
yields the density of condensate particles. 
%This rough approximation is compared with the ideal Bose gas in figure \ref{n0}. 
The critical temperature is increasing with increasing scattering length. Normalizing the critical temperature to unity we found no difference to the result of a selfconsistent calculation \cite{SG98}.

%\begin{figure}[h]
%\psfig{file=n0.eps,width=8cm}
%%\vspace*{-4ex}
%\caption{
%The temperature dependence of the fraction of condensate particles. The ideal Bose gas ($a_0=0$), $n_0/n=1-(T/T_c)^{3/2}$, is compared with the freezing model used here.}\label{n0}
%\end{figure}

The analysis so far has convinced us that we can find the BEC from the pole condition of the T-matrix. It is now interesting to note that this BEC pole at negative energies means also that there is a two-particle binding energy. We found the puzzling property that in the BEC regime the particles show a binding energy. 
We can calculate the pole condition of the T-matrix analytically in the zero-temperature limit, where the bose function in (\ref{fp}) vanishes and the condensate density $n_0$ is equal to the total one $n$. Introducing the condensate part of (\ref{fp}) in (\ref{J}) we obtain, assuming a quadratic dispersion,
\be
J\left(Q,\omega+\frac{Q^2}{4m}\right)&=&a_0\Theta(-\omega)\sqrt{-\frac{m\omega}{\hbar^2}}-ia_0\Theta(\omega)\sqrt{\frac{m\omega}{\hbar^2}}\nonumber\\
&+&{\Delta \over \omega -\frac{Q^2}{4m}}
\ee
with the characteristic energy
$
\Delta=8\pi n a_0 {\hbar^2 \over m}.
%\label{Delta}
$ The critical momentum follows as
\be
Q_c&=&2\sqrt{m \Delta}
\label{qc2}
\ee
and the binding energy $E_B$ can be obtaint from
\be
-1=J\left(Q,-E_B\!+\!\frac{Q^2}{4m}\right)=a_0\sqrt{\frac{mE_B}{\hbar^2}}\!-\!{\Delta \over E_B \!+\!\frac{Q^2}{4m}}.
%\nonumber\\
\ee
A comparison with the heuristic result (\ref{qc1}) shows a difference by a factor of $\sqrt{3}$ which is quite satisfying for the level of approximations. Both approximations, (\ref{qc1}) and (\ref{qc2}), are compared in figure~\ref{qct2}. For small scattering lengths the heuristic approximation (\ref{qc1}) overestimates the numerical value since it neglects the relative motion of the two particles. The dominant-condensate approximation (\ref{qc2}) underestimates the critical momentum since the correlations of normal particles are neglected.

An expansion of the binding energy around its root, i.e., the critical momentum, yields a linear dispersion $E_B(Q)\approx Q_c(Q-Q_c)/2m$ 
from which the Landau critical velocity for superfluidity \cite{PS04} of $V_C=Q_c/2m$ is obtained which also justifies our interpretation of $Q_c$ as the critical momentum of superfluidity.

\begin{figure}
\parbox[t]{9.1cm}{
\parbox[t]{4.5cm}{\psfig{file=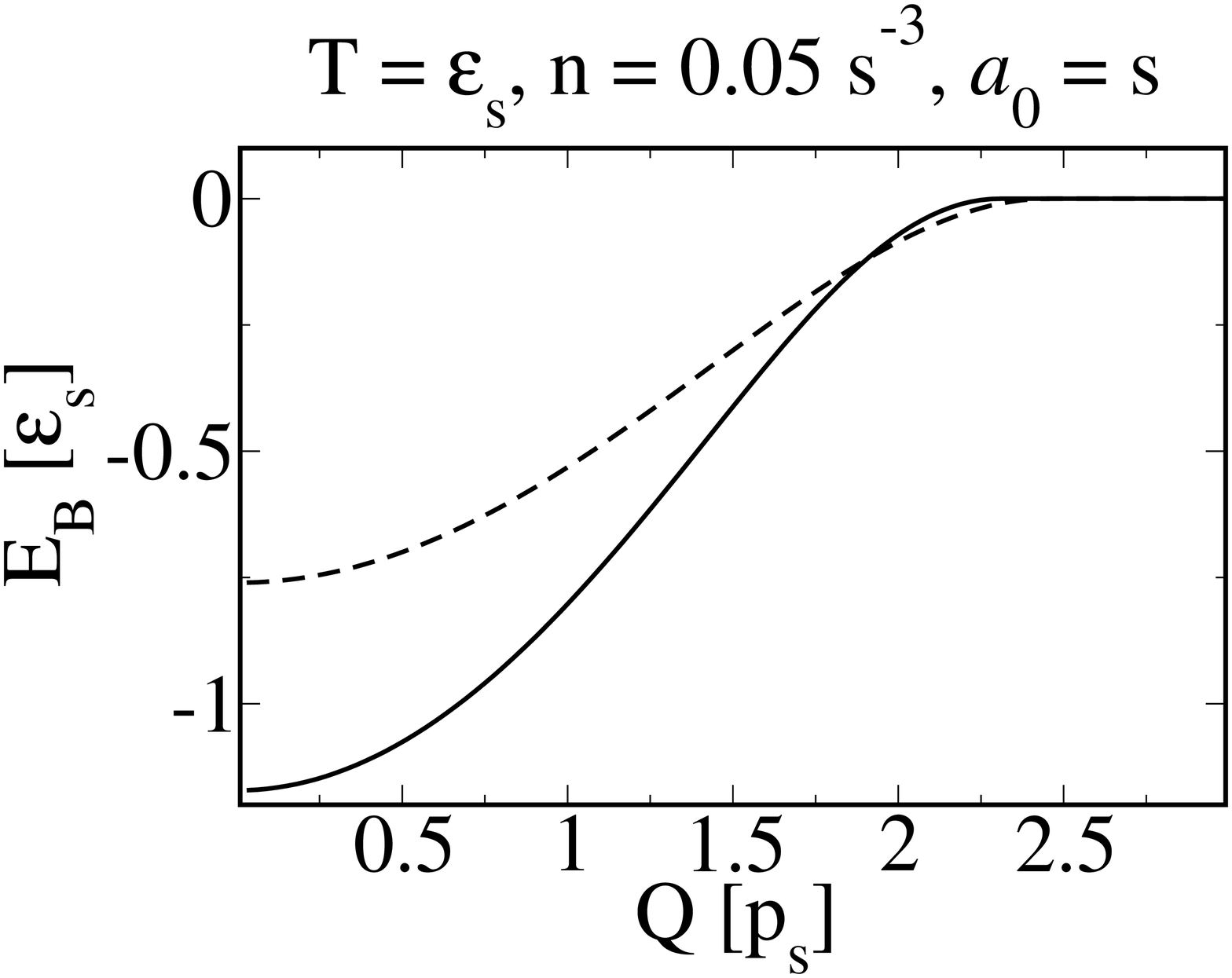,width=4.5cm}
}%\hspace*{-2ex}
\parbox[t]{4.5cm}{\psfig{file=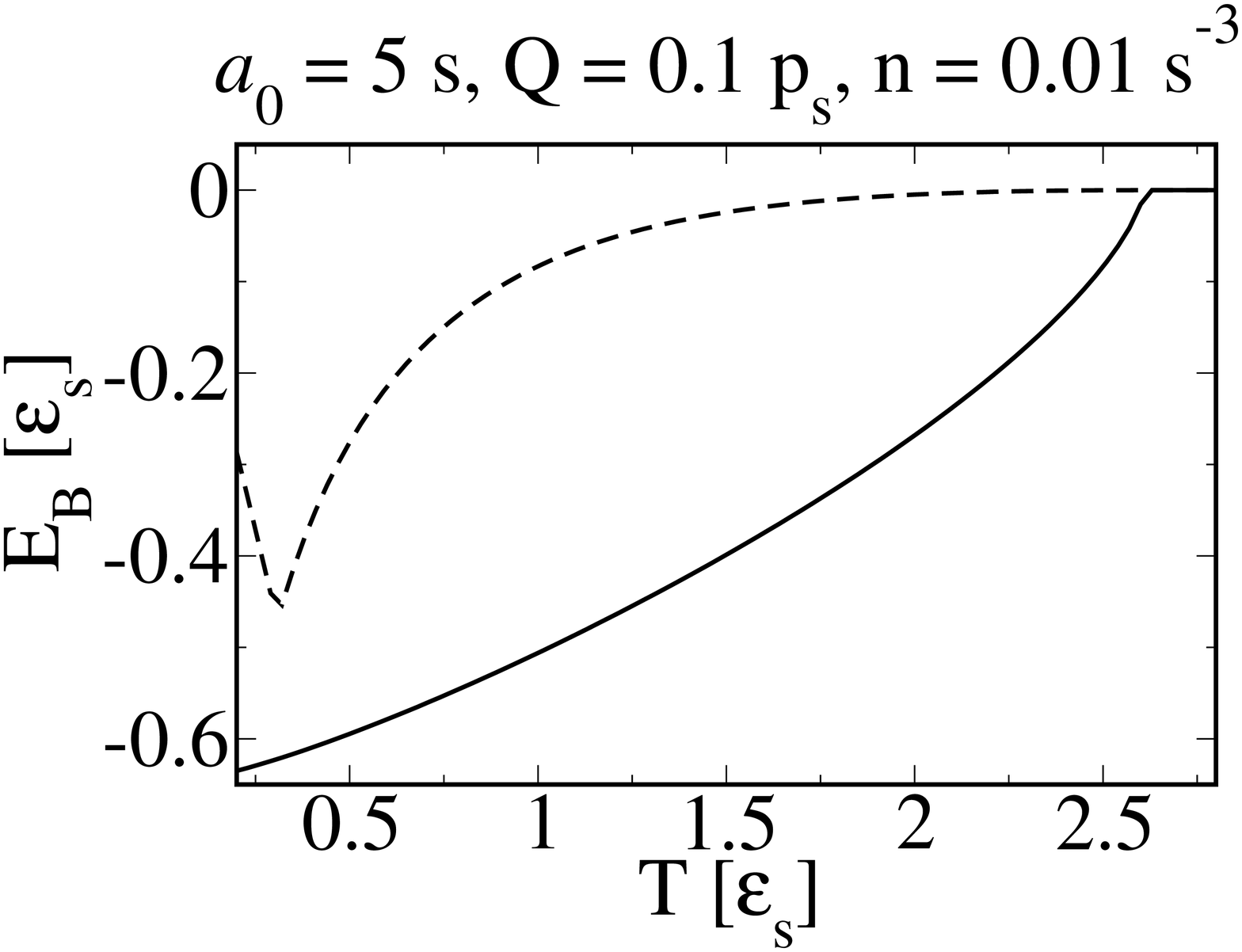,width=4.5cm}
}
}
\vspace*{-2ex}
\parbox[t]{9.1cm}{
\parbox[t]{4.5cm}{\psfig{file=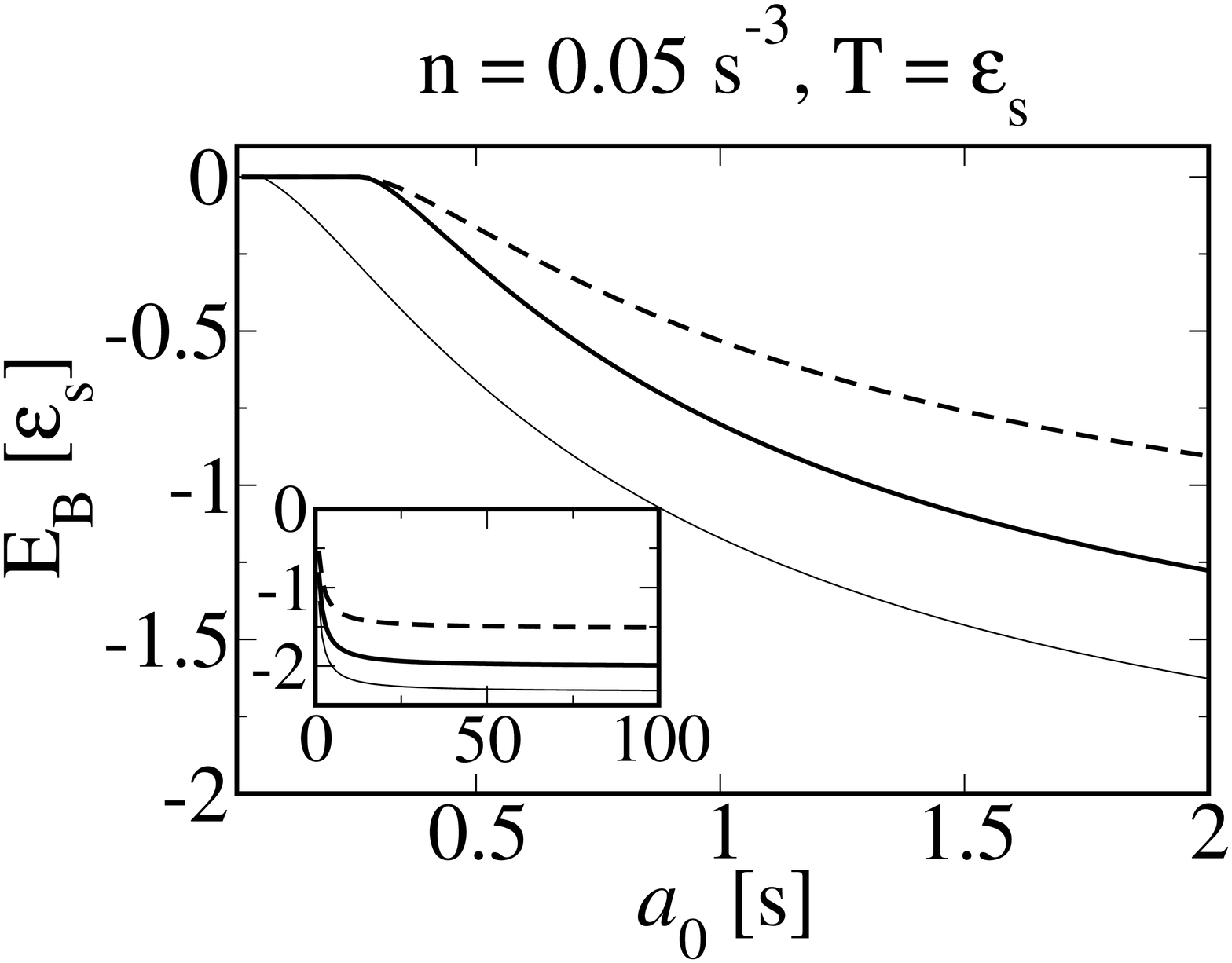,width=4.5cm}
}%\hspace*{-2ex}
\parbox[t]{4.5cm}{\psfig{file=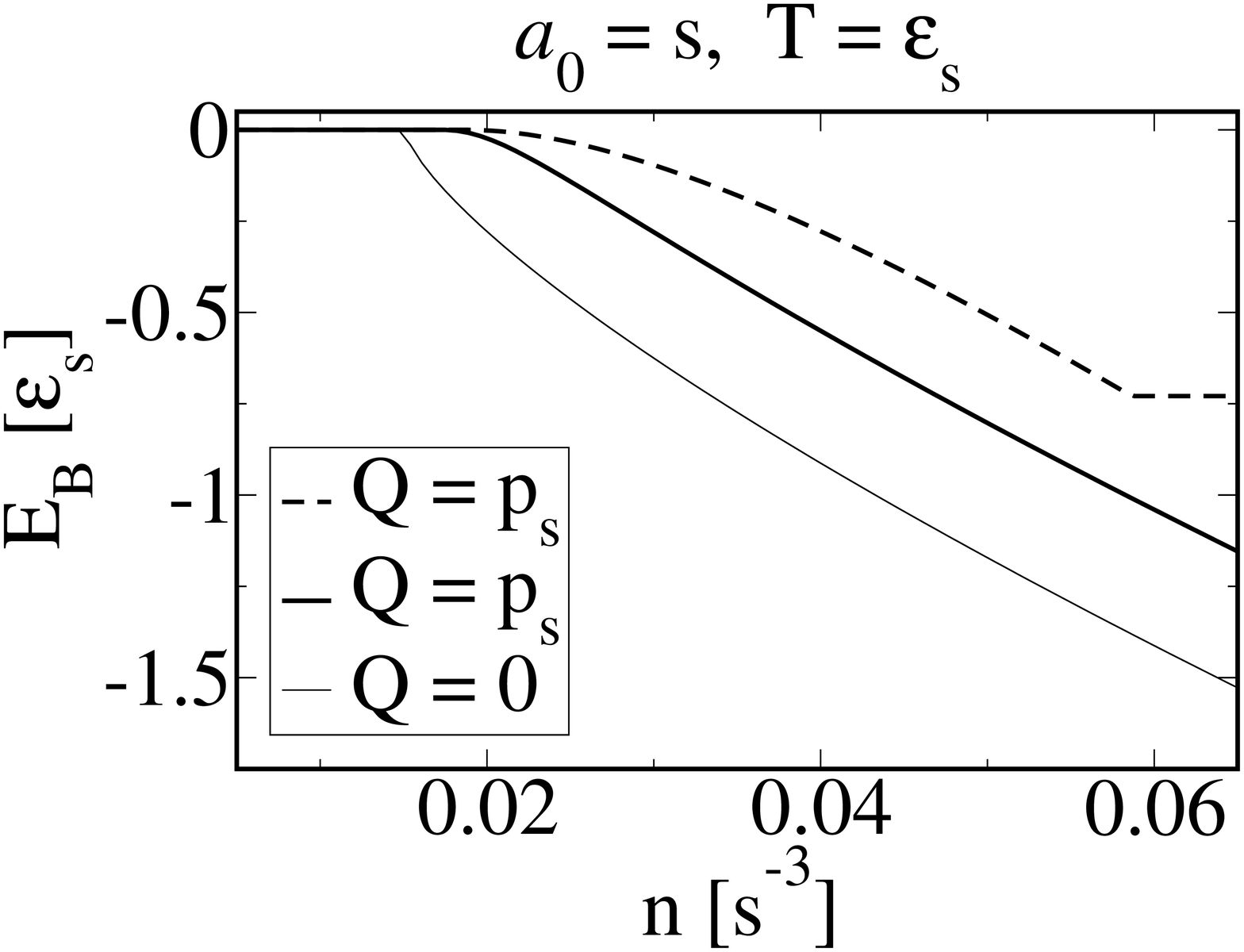,width=4.5cm}
}
}
\caption{The two-particle binding energy of the BEC relative to the continuum edge dependent on the different parameters keeping the other ones fixed. The broken line indicates the result without condensate density for finite momentum. The thin solid line is the result in the center-of-mass frame while the thick one is for finite momentum. The inset shows the strong coupling limit of the binding energy.}\label{eb}
\end{figure}

The appearance of the binding energy is connected with the critical density and temperature for BEC as one can see in figure \ref{eb}. The binding energy becomes larger  with increasing density or scattering length. For larger scattering length the binding energy saturates. Going beyond the T-matrix approximation
a nonlinear behavior is expected analogous to the critical temperature \cite{MMS07}. The binding energy decreases with increasing temperature or momentum which follows the reasoning concerning the behavior of the condensate. 

Summarizing we have analyzed the pole of the medium-dependent T-matrix and found the condition for the BEC from this pole. Besides the phase diagram this pole allows us to determine the critical momentum for the occurrence of superfluidity. As a new observation we found that the BEC is linked to the occurrence of a two-particle binding energy. We suggest that this could have a cosmological relevance concerning the dark energy problem since the energy balance of interacting Bose systems is changed by this two-particle binding energy.

This work was supported by the German DAAD and Czech research plan MSM 0021620834.

\appendix

\bibliography{sem3,bose,kmsr,kmsr1,kmsr2,kmsr3,kmsr4,kmsr5,kmsr6,kmsr7,delay2,spin,refer,delay3,gdr,chaos,sem1,sem2,short}

\end{document}